\documentclass[prb,reprint,superscriptaddress,aps]{revtex4-2}

\usepackage[normalem]{ulem}
\usepackage{graphicx}
\usepackage{bm}
\usepackage{epstopdf}
\usepackage{amsmath}
\usepackage{amssymb}
\usepackage[ansinew]{inputenc}
\usepackage{multirow}
\usepackage[urlcolor=blue,colorlinks=true,citecolor=blue,linkcolor=blue,
            bookmarks=false]{hyperref}
\usepackage[dvipsnames]{xcolor}
\usepackage[OT4]{fontenc}
\begin{document}
\title{Coupling between magnetic and thermodynamic properties in
$R$Rh$_2$Si$_2$ ($R$ = Dy, Ho)}

\author{H. Dawczak-D\k{e}bicki}
\affiliation{Max Planck Institute for Chemical Physics of Solids, D-01187
Dresden, Germany}

\author{K. Kliemt}
\affiliation{Kristall- und Materiallabor, Physikalisches Institut,
Goethe-Universit\"at Frankfurt, D-60438 Frankfurt/M, Germany}

\author{M. V. Ale Crivillero}
\affiliation{Max Planck Institute for Chemical Physics of Solids, D-01187
Dresden, Germany}

\author{R. K\"uchler}
\affiliation{Max Planck Institute for Chemical Physics of Solids, D-01187
Dresden, Germany}

\author{C. Krellner}
\affiliation{Kristall- und Materiallabor, Physikalisches Institut,
Goethe-Universit\"at Frankfurt, D-60438 Frankfurt/M, Germany}

\author{O. Stockert}
\affiliation{Max Planck Institute for Chemical Physics of Solids, D-01187
Dresden, Germany}

\author{S. Wirth}
\email[e-mail: ]{steffen.wirth@cpfs.mpg.de}
\affiliation{Max Planck Institute for Chemical Physics of Solids, D-01187
Dresden, Germany}

\date{\today}

\begin{abstract}
Single crystals of DyRh$_2$Si$_2$ and HoRh$_2$Si$_2$ were investigated by
thermal expansion and magnetostriction. The different types of magnetic order
can clearly be seen in these measurements, particularly the canting of the
moments away from the crystallographic $c$ direction below about 12~K and the
spin-flip for magnetic field applied along the $c$ direction. For
HoRh$_2$Si$_2$, an additional transition just below $T_{\rm N}$ is analyzed
by means of the Gr\"{u}neisen ratio and is likely caused by a change of the
magnetic structure. Our results nicely corroborate findings from other magnetic
and thermodynamic measurements on these materials and provide further evidence
suggesting the formation of magnetic domains.
\end{abstract}
\maketitle

\section{Introduction}
Materials crystallizing in the ThCr$_2$Si$_2$-type structure (space group
$I4/mmm$) exhibit a variety of interesting physical phenomena \cite{hof85}
including superconductivity \cite{si16,ste16,sha19}. More specifically, the
discovery of superconductivity in some rare-earth compounds of this family
\cite{ste79,pal85,mov96,mat98,sch16} provided enormous insight into, and
propelled, the field of heavy-fermion physics and beyond \cite{mat98,yua03,
bro08}. In particular, these compounds gave some valuable glue concerning the
pivotal impact of magnetism on unconventional superconductivity and quantum
criticality \cite{kei15}. In consequence, it is vital to deepen our insight
into the variety of magnetic properties and phenomena of these materials
\cite{lai22}.

Even within the rare-earth 122 series the magnetic properties vary widely. Ce-
and Yb-based materials often exhibit non-integer valencies of the rare-earth
($R$) and are discussed in terms of RKKY interaction
(Ruderman-Kittel-Kasuya-Yosida, \cite{rud54,kas56,yos57}) mediated via a
polarization of the conduction electrons. This interaction can compete with the
Kondo effect, an on-site screening of the 4$f$ moments by the conduction
electrons \cite{ste16}. For stable, trivalent $R$ like Nd, Gd, Tb, Ho or Er,
the RKKY interaction results in local moment antiferromagnetic (AFM) order,
often with a simple propagation vector of \textbf{Q} = (001) and ferromagnetic
ordering within the plane perpendicular to (001) \cite{sla83,fel84,mel84,szy84,
mel98,hos09,kli17,sic18}. In many compounds the local moments align along the
crystallographic $c$ direction, while for SmRh$_2$Si$_2$, GdRh$_2$Si$_2$ and
GdIr$_2$Si$_2$ an orientation in the $ab$ plane is reported \cite{kli17,sic18,
kli19} which, in case of GdRh$_2$Si$_2$, can even be temperature dependent
\cite{win20}. Here it should be noted that in the $R$-based compounds the
spin-orbit coupling is generally larger compared to crystal field effects
\cite{rin11}. In addition, anisotropic exchange was discussed for
TbRh$_2$Si$_2$ \cite{che85}. It should also be noted that $d$-electrons of the
transition elements (e.g. Rh, Ru) contribute very little to the total magnetic
moment; the value of $\sim$0.002 $\mu_{\rm B}$ for Rh in DyRh$_2$Si$_2$ is too
small to be detected in neutron diffraction \cite{sla83,mel84}. In case of
$R =$ Gd, there can also be a small contribution from Gd 5$d$ electrons,
$\sim$0.28 $\mu_{\rm B}$ in GdRh$_2$Si$_2$ \cite{czj89}.

Exceptions to the above-mentioned magnetic configuration are DyRh$_2$Si$_2$
and HoRh$_2$Si$_2$ where the magnetic moments were found to be canted away from
the crystallographic $c$ axis by neutron diffraction \cite{sla83,mel84}. For
both compounds, the magnetic properties and specific heat measurements were
analyzed in terms of a mean field model \cite{tak87,tom89,tak92,kli23}. Yet,
based on magnetic susceptibility and specific heat measurements a so-called
component-separated magnetic transition, stemming from multiple interactions,
was suggested for HoRh$_2$Si$_2$ \cite{shi11}, in analogy to the tetragonal
compound TbCoGa$_5$ \cite{san09}. To gain further insight, measurements of
thermodynamic properties are called for, in particular such that allow to
provide information along different crystallographic directions of the sample.
Therefore, we conducted measurements of thermal expansion and magnetostriction
on single crystalline DyRh$_2$Si$_2$ and HoRh$_2$Si$_2$. To allow for
comparison to data from literature, magnetic susceptibility was also measured
on the very same samples.

\section{Experimental}
The single crystals of DyRh$_2$Si$_2$ and HoRh$_2$Si$_2$ used in this study
were grown from In flux employing a modified Bridgman technique; details of the
growth procedure were provided in Ref.\ \cite{kli20}. X-ray diffraction on
powdered single crystals was conducted (using copper K$_{\alpha}$ radiation in
a Bruker D8 diffractometer) to confirm the crystallographic structure and
quality of the samples. The crystallographic orientation of the single crystals
was determined by Laue diffraction. The samples typically grew in a
platelet-like shape with the crystallographic $c$ direction (the long axis of
the tetragonal unit cell) along the thin sample dimension. In most cases, the
other sample edges were parallel to the $\langle$110$\rangle$ crystallographic
directions.

The thermal expansion and magnetostriction measurements were performed using a
dilatometer cell as described in Ref.\ \cite{kue17,kue23}. The measurements
were conducted in a Physical Property Measurement System (PPMS) by Quantum
Design, Inc.\ with maximum magnetic field of 9 T applied parallel to the sample
dilatation direction investigated. Here, special attention was paid to minimize
electrical noise \cite{kue23}. Whenever possible, identical samples were used
for measurements along different crystallographic directions. However, in some
cases samples were chosen according to their specific shape in order to
optimize the dilatometer signal for the crystallographic direction to be
measured and to assist sample mounting. Typical sample dimensions were 1--2~mm
in the $ab$ plane and up to 0.6 mm along the $c$ direction. Data of the thermal
expansion were taken upon warming the sample (if not stated otherwise) and
repeated at least once for comparison. Between cycles of magnetostriction
measurements (i.e.\ measurements at constant temperature) the sample was warmed
up to at least 80 K, i.e.\ into the paramagnetic state well above the N\'{e}el
temperatures $T_{\rm N} \approx$ 55~K for DyRh$_2$Si$_2$ and $T_{\rm N}
\approx$ 29~K for HoRh$_2$Si$_2$. We note that different samples gave somewhat
different results for the thermal expansion and magnetostriction, particularly
in the temperature range around $T_1$ and for small magnetic fields,
respectively, as discussed below. The observed temperatures and magnetic fields
of the transitions, however, reproduced very well.

Measurements of the magnetic dc susceptibility were conducted in a Magnetic
Property Measurement System (MPMS3 by Quantum Design, Inc.) using the same
samples as for the thermal expansion and magnetostriction measurements. For
these measurements, a magnetic field of 25 Oe (corrected for the remnant field
of the superconducting magnet as determined by a Pd reference) was applied.
The PPMS, equipped with a calorimeter that utilizes a quasi-adiabatic thermal
relaxation technique, was also used for measurements of the heat capacitance.
For the investigation of possible first-order transitions, we used a
single-slope analysis of the measured heat pulses, as described in Ref.\
\cite{las03}.

\section{Results}
\subsection{HoRh$_2$Si$_2$}
\begin{figure}[t]
\centering
\includegraphics[width=8.6cm]{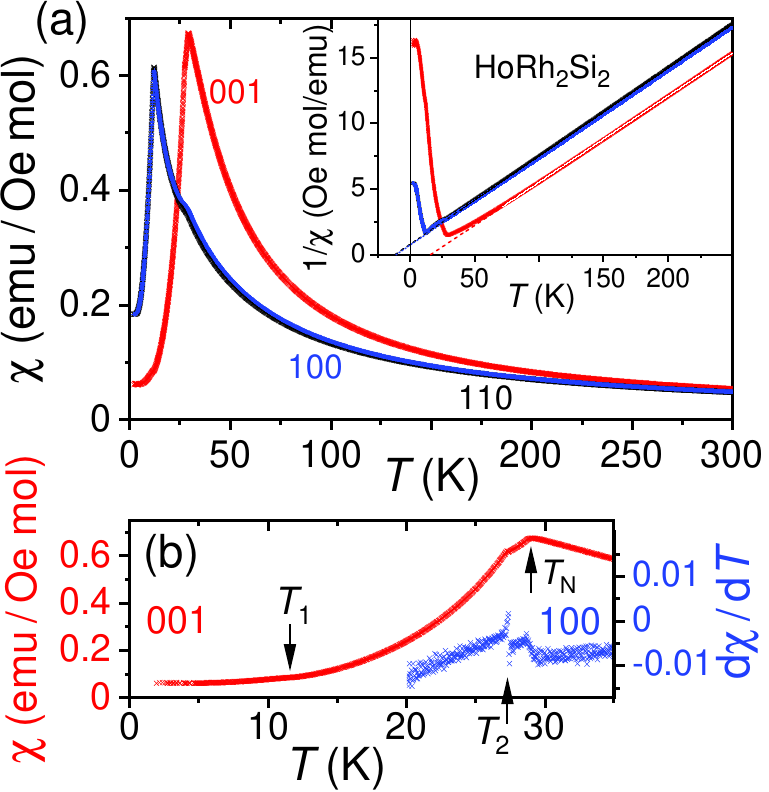}
\caption{(a) dc susceptibility $\chi$ measured for HoRh$_2$Si$_2$ along three
different crystallographic directions in a field of 25 Oe. The inset shows the
inverse of the susceptibilities. The dashed/dotted lines represent
extrapolations of the linear fits to $1/\chi$. (b) Zoom into the low-$T$ data
of $\chi (T)$ for the (001) direction (red). The right scale visualizes d$\chi
(T)$/d$T$ (blue, in units of emu\,Oe$^{-1}$mol$^{-1}$K$^{-1}$) within 20~K
$\le T \le$ 35~K along (100) direction. Arrows mark the small hump at $T_1 =$
11.7~K and features at $T_2 =$ 27.3~K and $T_{\rm N} =$ 29.0~K, respectively.
Note the uniform color code for the different directions in all plots.}
\label{suscep}  \end{figure}

The results of dc susceptibility $\chi$ measurements for HoRh$_2$Si$_2$ along
different crystallographic directions are presented in Fig.\ \ref{suscep}(a).
The N\'{e}el temperature of $T_{\rm N} = 29.0$~K is clearly observed when
measured along the (001) direction. Additional small humps can be recognized
upon zoom into the low-temperature data, red data in Fig.\ \ref{suscep}(b),
at $T_2 =$ 27.3~K and $T_1=$ 11.7~K (the latter can also be seen in the
$1/\chi$-plot, inset of Fig.\ \ref{suscep}(a)). Along the (100) and (110)
direction, $\chi (T)$ peaks sharply at 11.9~K, i.e. at the temperature of the
small hump in $\chi (T)$ along (001), while small kinks are seen at
$T_{\rm N}$. Only upon taking the derivatives d$\chi (T)$/d$T$, these small
kinks separate into two features at $\sim$27.3~K and $\sim$29.0~K as shown for
the (100) direction in Fig.\ \ref{suscep}(b), blue data and right scale. All
$\chi (T)$-data nicely follow a Curie-Weiss law in the paramagnetic regime as
is obvious from plots of $1/\chi$ in the inset of Fig.\ \ref{suscep}(a). The
fits yield effective moments of $\mu_{\rm eff} \approx 10.9\, \mu_{\rm B}$
which is slightly larger than the expected value of $10.61\, \mu_{\rm B}$ for
Ho$^{3+}$ (where $\mu_{\rm B}$ is the Bohr magneton). This may be attributed to
the small magnetic field applied during the susceptibility measurements and the
resulting impact of the remnant field. The obtained Weiss temperatures are
$\theta_c \sim$ 16~K along the $c$ direction, indicative of dominating
ferromagnetic interactions, and $\theta_{ab} \sim -11$~K within the $ab$ plane
(dominating antiferromagnetic interactions). All results are in good agreement
with reported ones \cite{sla83,fel83,jaw02,shi11,shi12,kli20,usa23} and confirm
the magnetic properties as outlined in the introduction. In addition, the
observation of two transitions near $T_{\rm N}$ along all measured directions
indicates high sample quality.

\begin{figure}[t]
\centering
\includegraphics[width=8.4cm]{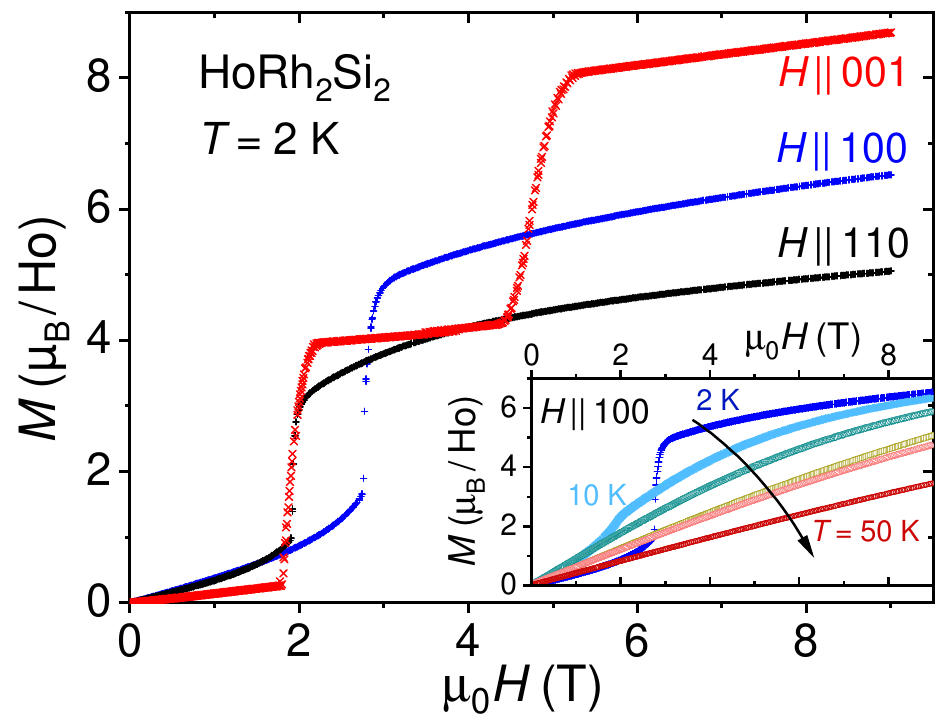}
\caption{Field dependence of magnetization $M$ measured at 2 K and for magnetic
fields $H$ applied along different directions (as indicated). The inset shows
the temperature evolution of $M(H)$ for $H \parallel (100)$ at $T =$ 2, 10, 15,
26, 28 and 50 K.}  \label{magn}
\end{figure}

The magnetic field ($H$) dependence of magnetization $M$ measured at $T = 2$~K
and with $H$ applied along different crystallographic directions is presented
in Fig.\ \ref{magn}. Even for $\vec{H} \parallel$ (001) the expected saturation
magnetic moment of Ho$^{3+}$, $g \mu_{\rm B} J = 10\,\mu_{\rm B}$, is not
observed for $\mu_0 H \leq$ 9~T. However, the two-step magnetization increase
by about 4~$\mu_{\rm B}$ per step agrees well with the reported canting of the
magnetic moments away from the crystallographic $c$ direction by $\sim\!
28^{\circ}$ at $T =$ 4.2~K \cite{sla83}. The two-step magnetization increase
itself is a consequence of the propagation vector $\vec{k} = (0,0,1)$ and a
change in magnetic configuration with increasing $H$ from an AFM +$-$+$-$ state
at $\mu_0 H \lesssim$ 1.8~T to +++$-$, and finally a tilted, field-polarized
++++ state beyond 5~T \cite{sla83,shi12}.

For $\vec{H} \parallel$ (100), the basal plane component of the magnetic
moments within the different magnetic domains is initially rotated toward the
field direction, and then flips to the (100) direction parallel to $\vec{H}$
near 2.8~T. This is suggested by the magnetization value of approximately
4.9~$\mu_{\rm B}$ at fields just beyond the flip which is only slightly larger
than the expected value for a canting angle of $28^{\circ}$. At $T =$ 10~K,
this canting angle is markedly smaller and hence, a considerably smaller
magnetization value is observed beyond the flip (see inset of Fig.\
\ref{magn}). For $T > T_1$, i.e. without canting, such a flip of the
magnetization is neither expected nor observed. The magnetization behaviour for
$\vec{H} \parallel$ (110) is qualitatively very similar to the observations for
$\vec{H} \parallel$ (100); the smaller magnetization values for large fields,
however, indicates the (110) direction to be magnetically harder compared to
the (100) direction. We note that our $M(H)$-behavior for in-plane applied
fields differs from the reported one \cite{shi12}.

\begin{figure}[t]
\centering
\includegraphics[width=8.0cm]{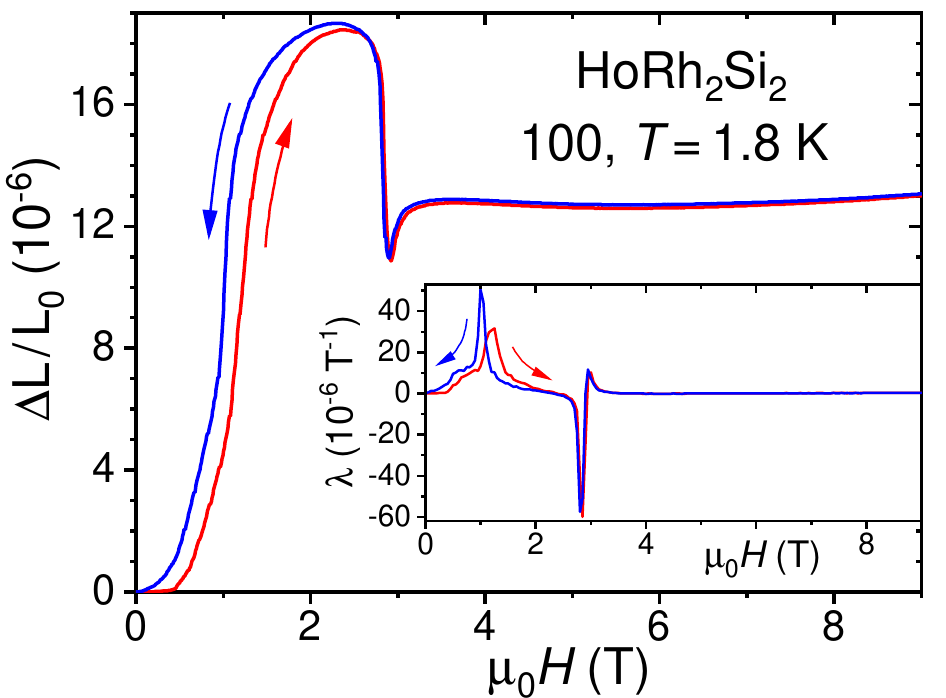}
\caption{Magnetostriction $\Delta {\rm L}(H) / $L$_0$ of HoRh$_2$Si$_2$
measured at $T =$ 1.8~K with magnetic fields applied along the (100)
crystallographic direction while sweeping the field up (red) and down (blue).
Inset: Magnetostriction coefficient $\lambda =$ (1/L$_0)\, \partial {\rm L} /
\partial H$.}   \label{mag-stric}
\end{figure}

Large magnetostriction is commonly observed in rare earth-containing compounds
due to their orbital magnetism. Within a quadrupole approximation, the 4$f$
electron densities of Ho$^{3+}$ and Dy$^{3+}$ retain an oblate shape
\cite{rin11}. In Fig.\ \ref{mag-stric}, the magnetostriction $\Delta {\rm L}(H)
/ $L$_0$ [where L$_0(T)$ = L($T,H =$0)] and its coefficient $\lambda =$
(1/L$_0)\,\partial {\rm L} / \partial H$ at $T =$ 1.8~K is presented for
$\vec{H} \parallel$ (100). These data are in excellent concert with the
$M(H)$-data of Fig.\ \ref{magn}. We observe a large increase of L$(H)$ upon
rotation of the in-plane component of magnetization, while the flip at $\mu_0 H
\approx$ 2.8~T results in a sudden drop of L$(H)$. As expected from this
scenario, there is only very little length change for fields above 3~T. The
discrepancy between the magnetostriction measured during up-sweep (red lines in
Fig.\ \ref{mag-stric}) and down-sweep (blue) is in line with a scenario
involving different magnetic domains in this tetragonal material.

\begin{figure}[t]
\centering
\includegraphics[width=8.4cm]{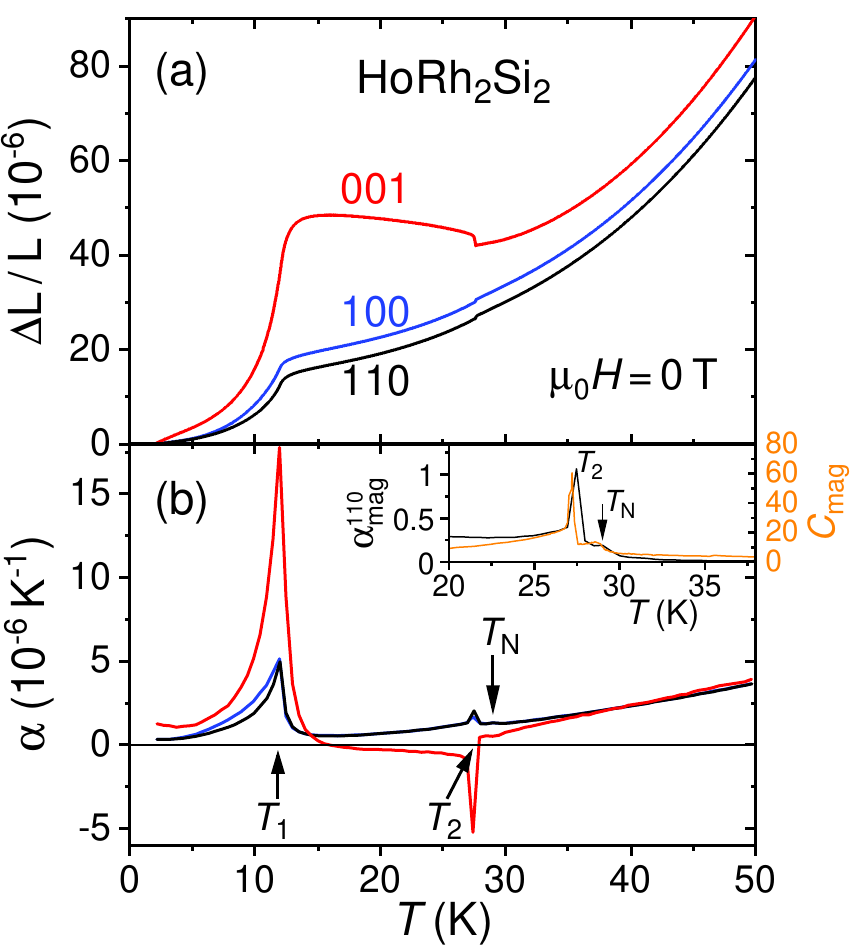}
\caption{(a) Relative length change $\Delta {\rm L}(T) / $L and (b) thermal
expansion coefficient $\alpha (T)$ of HoRh$_2$Si$_2$ measured along different
crystallographic directions in the temperature range 2~K $\le T \le$ 50~K.
Inset: Comparison between $\alpha^{110}_{\rm mag}$, the magnetic contribution
to $\alpha_{110}$ along the (110) direction, and $C_{\rm mag}$ from Fig.\
\ref{specheat}(a). Here, the units are: $\alpha^{110}_{\rm mag}$ in
10$^{-6}\,$K$^{-1}$ and $C_{\rm mag}$ in J$\,$mol$^{-1}\,$K$^{-1}$.}
\label{Ho-TE}  \end{figure}

Figure \ref{Ho-TE} exhibits the relative length changes $\Delta {\rm L}_i (T)
/ {\rm L}_i$ and the uniaxial thermal expansion coefficients $\alpha_i = (1 /
{\rm L}_i) (d {\rm L}_i /d T)$ for HoRh$_2$Si$_2$. Here, the index $i$ denotes
measurements along the different crystallographic directions (100), (110) and
(001). Clearly, strong maxima in $\alpha_i (T)$ are observed at $T_1 =$ 11.8~K
for all directions, a temperature which agrees well with the feature observed
in $\chi (T)$. Additional peaks are observed for all directions upon
approaching $T_{\rm N}$. Importantly, equal-area constructions for $\Delta
{\rm L}_i (T) / {\rm L}_i$ yield temperatures of the jumps of 27.6~K for the
(100) and (110) directions, and 27.5~K for (001), i.e. very close to $T_2$. In
contrast, only tiny variations of $\alpha (T)$ are present at $T_{\rm N}$, see
respective arrow in Fig.\ \ref{Ho-TE}(b). We note that magnetization
measurements along (001) only showed a small cusp at $T_2 =$ 27.3~K
\cite{shi12} while the magnetic specific heat peaked dramatically at this
temperature (cf.\ discussion below and \cite{shi11}), the latter very similar
to our thermal expansion results. Clearly, our thermal expansion measurements
are less sensitive to the onset of magnetic order at $T_{\rm N}$ and point to
a mechanism operating at $T_2$ which is different from the antiferromagnetic
ordering at $T_{\rm N}$. Moreover, at $T_2$ the $\Delta {\rm L}_i (T) /
{\rm L}_i$ jump occurs in opposite directions: $\Delta {\rm L}_{100} (T) /
{\rm L}_{100}$ and $\Delta {\rm L}_{110} (T) / {\rm L}_{110}$ expand by about
$0.45 \cdot 10^{-6}$ (obtained from equal-area constructions around the jumps)
but HoRh$_2$Si$_2$ contracts along $c$ by $\Delta {\rm L}_{001} (T) /
{\rm L}_{001} \approx -2.1 \cdot 10^{-6}$ and hence, the volume shrinks upon
warming the sample through $T_2$. This anisotropic thermal expansion is in
line with the reported increase of the $c/a$ ratio upon cooling from room
temperature to 4.2 K \cite{sla83}. There is, however, no indication for any
discontinuous change in the lattice constants as, e.g., observed for some
ThCr$_2$Si$_2$-type phosphides \cite{huh97}.

We note that our repeated measurements of thermal expansion (also on different
samples) all agree qualitatively, but vary quantitatively for the (001)
direction, particularly within the range $T_1 \lesssim T < T_2$. We speculate
that domain formation may play a role in generating such quantitative
differences. This is supported by the fact that the transition temperatures
$T_1$ and $T_2$ themselves agree nicely for all measurements conducted.

\begin{figure}[t]
\centering
\includegraphics[width=8.6cm]{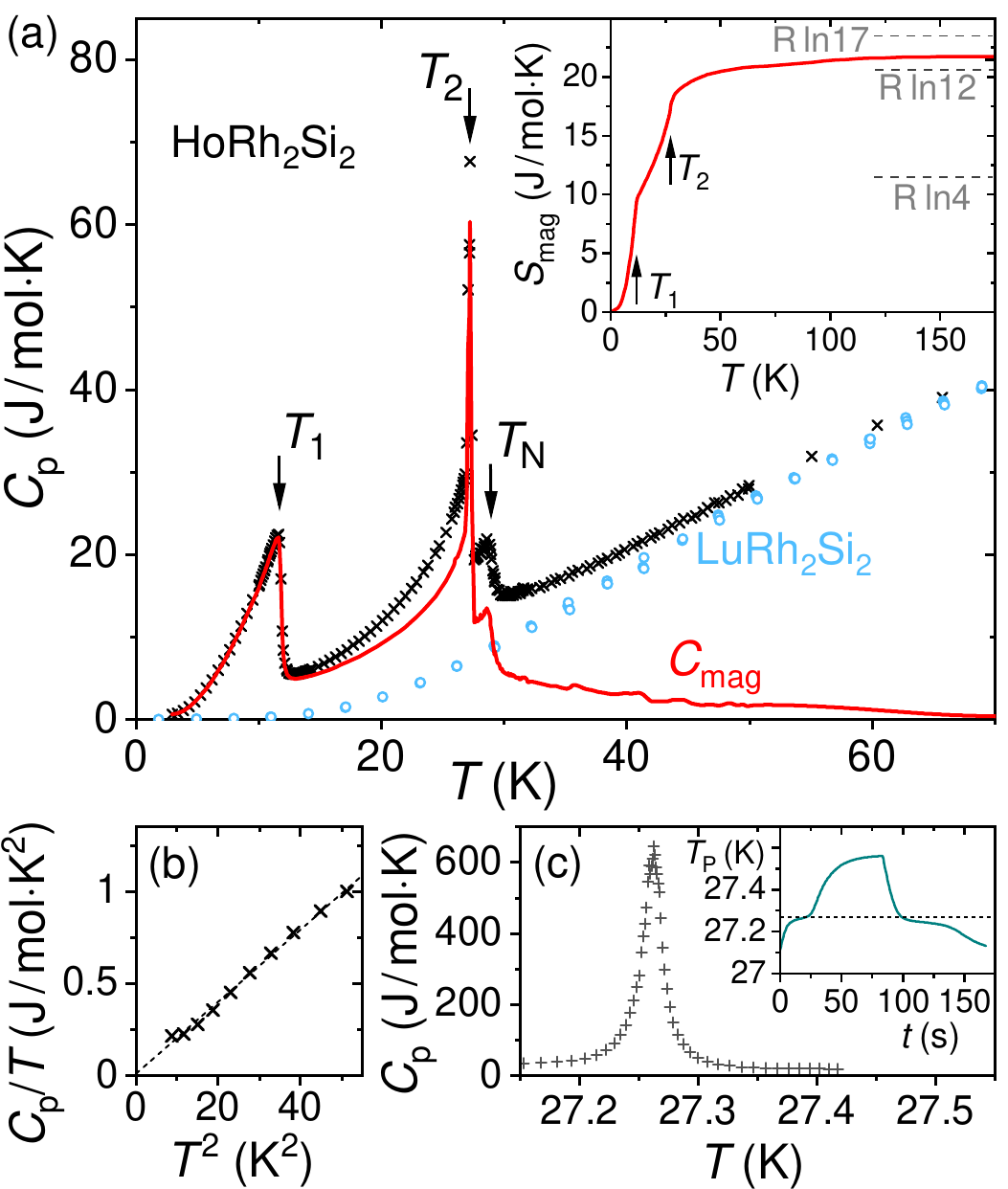}
\caption{(a) Temperature dependence of the specific heat $C_p(T)$ of
HoRh$_2$Si$_2$ and the non-magnetic reference compound LuRh$_2$Si$_2$
\cite{fer07}. The magnetic contribution $C_{\rm mag}$ (red line) to $C_p(T)$ of
HoRh$_2$Si$_2$ is obtained as described in the text. Inset: Magnetic entropy
$S_{\rm mag}$ as calculated from $C_{\rm mag}$. (b) $C_p(T)/T$ vs. $T^2$
at lowest temperatures along with a linear fit. (c) High-resolution $C_p(T)$
measurement near $T_2$ (see text). Inset: Arrest of the measurement platform
temperature $T_{\rm P}$ at the sample's first order transition at $T_2$.}
\label{specheat}
\end{figure}

Given this unusual behavior of the thermal expansion, we performed measurements
of the specific heat $C_p(T)$ on HoRh$_2$Si$_2$. Figure \ref{specheat}(a)
presents the $C_p(T)$ data up to $T =$ 70~K albeit measurements were conducted
within 3~K $\le T \le$ 200~K. As the main result and in support of our thermal
expansion measurements, the largest peaks in $C_p(T)$ of HoRh$_2$Si$_2$ are
observed at $T =$ 11.6~K $\approx T_1$ and $T =$ 27.2~K $\approx T_2$, while
only a much less-pronounced shoulder is seen at $T_{\rm N}$ \cite{sek87}. At
lowest $T$, a Sommerfeld coefficient of $\gamma \approx$ 10.5 mJ$\,$mol$^{-1}
\,$K$^{-2}$ is estimated from Fig.\ \ref{specheat}(b), a value close to the
one (9.6~mJ$\,$mol$^{-1} \,$K$^{-2}$) obtained for trivalent Eu in
isostructural EuCo$_2$Si$_2$ \cite{sei19}. A temperature dependence of $C_p
\propto T^3$ is consistent with spin-wave excitations in an antiferromagnet
\cite{las08}. Nuclear contributions of Ho (with nuclear spin $I = 7/2$) to
$C_p(T)$ are expected to be negligibly small at the temperatures of interest
here \cite{kum16,sto20} while phonon contribution only add minimally to the
$T^3$-dependence of $C_p$. In order to evaluate the magnetic contribution
$C_{\rm mag}$ to the total specific heat of HoRh$_2$Si$_2$, the isostructural
compound LuRh$_2$Si$_2$ was taken as a non-magnetic reference
\cite{fer07}. However, the mass per formula unit of LuRh$_2$Si$_2$ exceeds the
one of HoRh$_2$Si$_2$ by approximately 2.4\%. Hence, for a more accurate
estimate of the phonon contribution in HoRh$_2$Si$_2$, $C_p(T)$ of
LuRh$_2$Si$_2$ was rescaled to account for its heavier atomic mass, as outlined
in \cite{tar03,sto20}. The so-determined $C_{\rm mag}$ of HoRh$_2$Si$_2$ is
shown by a red line in Fig.\ \ref{specheat}(a), and an estimate of the
resulting magnetic entropy $S_{\rm mag}(T) =\int_0^T (C_{\rm mag}(T')/ T') dT'$
is presented in the inset. Note that even at highest $T \gg T_{\rm N}$ the
magnetic entropy reaches only about 88\% of the expected value of R$\,\ln 17$
for Ho$^{3+}$, an observation which complicates an assessment of the associated
multiplet states. Nonetheless, one may speculate from $S_{\rm mag}(T_1)
\lesssim {\rm R}\,\ln 4$ that two doublets or one doublet and two singlets are
involved below $T_1$. In fact, if we use the crystalline electric field (CEF)
parameters as provided in \cite{usa23} we find a quasi-quartet ground state
made up of one doublet and two singlets within an energy range of less than
0.5~K. We note that Ho$^{3+}$ is a non-Kramers ion with 4 doublets and 9
singlets in tetragonal symmetry \cite{wal84}; a non-magnetic singlet ground
state is not evident. From the nice $T^3$-dependence of $C_p(T)$ ungapped
antiferromagnetic spin waves are expected; hence, the proposed quasi-quartet
ground state can be rationalized. Further, few of the singlets may reside high
up in energy which may explain the fact that the observed magnetization at 9~T
is distinctly smaller than the expected saturation magnetization (see
discussion above and Fig.\ \ref{magn}) as well as the ``missing'' entropy
$S_{\rm mag} <$ R$\,\ln 17$. We note that also in case of HoIr$_2$Si$_2$ a
saturation value of $S_{\rm mag} \sim$ R$\,\ln 12$ was reported \cite{kli18}
suggesting a common origin for finding a reduced $S_{\rm mag}$ at high
temperatures in both compounds.

Having both $\alpha_i (T)$ and $C_p(T)$ at hand, the Gr\"{u}neisen ratio
can be evaluated. For a single, dominating contribution to the entropy $S$ with
characteristic energy scale $T_j$, the Gr\"{u}neisen ratio $\Gamma_j$ is
expected to be independent of temperature \cite{gru08,geg16,dre20}.
Experimentally, this can be verified by analyzing the ratio $\alpha_i / C_p$
\cite{kli06}. In our case, we focus on the magnetic contributions to $\alpha_i$
and $C_p$. In order to estimate the phonon contribution to $\alpha_i$ we make
use of this contribution to $C_p$ as described above and scale it to $\alpha_i
(T\,=\, 70\: {\rm K})$. After subtraction, the resulting magnetic part
$\alpha_{\rm mag}^{110}$ is presented for the direction (110) in the inset to
Fig.\ \ref{Ho-TE}(b) along with $C_{\rm mag}$ [from Fig.\ \ref{specheat}(a)]
within 20~K $\le T \le$ 38~K for direct comparison. Clearly, the two quantities
scale reasonably well with $\alpha_{\rm mag}^{110} / C_{\rm mag} \sim 1.6
\cdot 10^{-8}$ mol/J. Similarly good agreement is found for the (100) direction
with $\alpha_{\rm mag}^{100} / C_{\rm mag} \sim 1.2 \cdot 10^{-8}$ mol/J, while
for (001) the agreement is not quite as nice. All this may indicate a common
magnetic origin of the transitions at $T_{\rm N}$ and $T_2$. The transition at
$T_1$ appears to be separate because of the much larger values of $\alpha_{\rm
mag}^i (T_1)$ compared to $\alpha_{\rm mag}^i (T_2)$ while the opposite holds
for $C_{\rm mag}$. Because of the extremely sharp peaks of $C_p$ and $\alpha_i$
at $T_2$ we find the population (or depopulation) of CEF levels as the cause of
this transition unlikely. Rather, we speculate that a change of the magnetic
structure takes place at $T_2$. The above described simple AFM structure with
$\vec{k} = (0,0,1)$ was established for $T \le 27 (\pm 1)$~K \cite{sla83}. This
leaves the possibility of a different magnetic structure within
$T_2 \le T \le T_{\rm N}$. In fact, Ho itself displays several magnetic
structures, including a helical one \cite{koe66}. An incommensurate magnetic
structure for $T_2 \le T \le T_{\rm N}$ was suggested in \cite{shi11}, yet
without showing data. In addition, a change from an incommensurate ordering
vector just below $T_{\rm N}$ to a commensurate one at lower $T$ was reported
for HoMn$_2$O$_5$ \cite{bla05} and \mbox{HoSbTe} \cite{plo22}, a similar
sequence was observed for HoNi$_2$B$_2$C \cite{gol94}. Therefore, a change of
the magnetic structure at $T_2$ is certainly possible but awaits confirmation,
e.g.\ by neutron scattering.

Beyond that, one may consider the impact of the magnetoelastic coupling on the
structure of HoRh$_2$Si$_2$. For instance, in some tetragonal rare-earth nickel
borocarbides a magnetostriction-induced orthorhombic lattice distortion was
observed \cite{det97,kre01,tof18}. Here, the lattice distortion (expressed as
the relative difference of the orthorhombic lattice parameters $a$ and $b$) was
reported to be proportional to the squared ordered magnetic moment around
$T_{\rm N}$ \cite{kre01}. In case of HoRh$_2$Si$_2$, one may then speculate
that the increase of the ordered magnetic moment upon cooling through
$T_{\rm N}$ may also, via magnetoelastic coupling, induce a structural phase
transition at $T_2$. Indeed, detailed measurements of $C_p (T)$ according
to the recipe outlined in Ref.\ \cite{las03} exhibit a very sharp peak at
$T_2$, Fig.\ \ref{specheat}(c). In addition, the measurement platform
temperature $T_{\rm P}$ arrested close to $T_2$ due to the sample's latent
heat \cite{las03}, inset to Fig.\ \ref{specheat}(c). Both observations clearly
indicate a first order transition taking place at $T_2$ supporting the
aforementioned scenario. However, high-resolution structural investigations in
this $T$-range are called for to substantiate such a speculation \cite{sek87}.

\subsection{DyRh$_2$Si$_2$}
A detailed description of the magnetic properties and the specific heat of our
DyRh$_2$Si$_2$ single crystals has been provided very recently \cite{kli23}
and hence, we will focus here on the thermal expansion and magnetostrictive
measurements.

\begin{figure}[t]
\centering
\includegraphics[width=8.4cm]{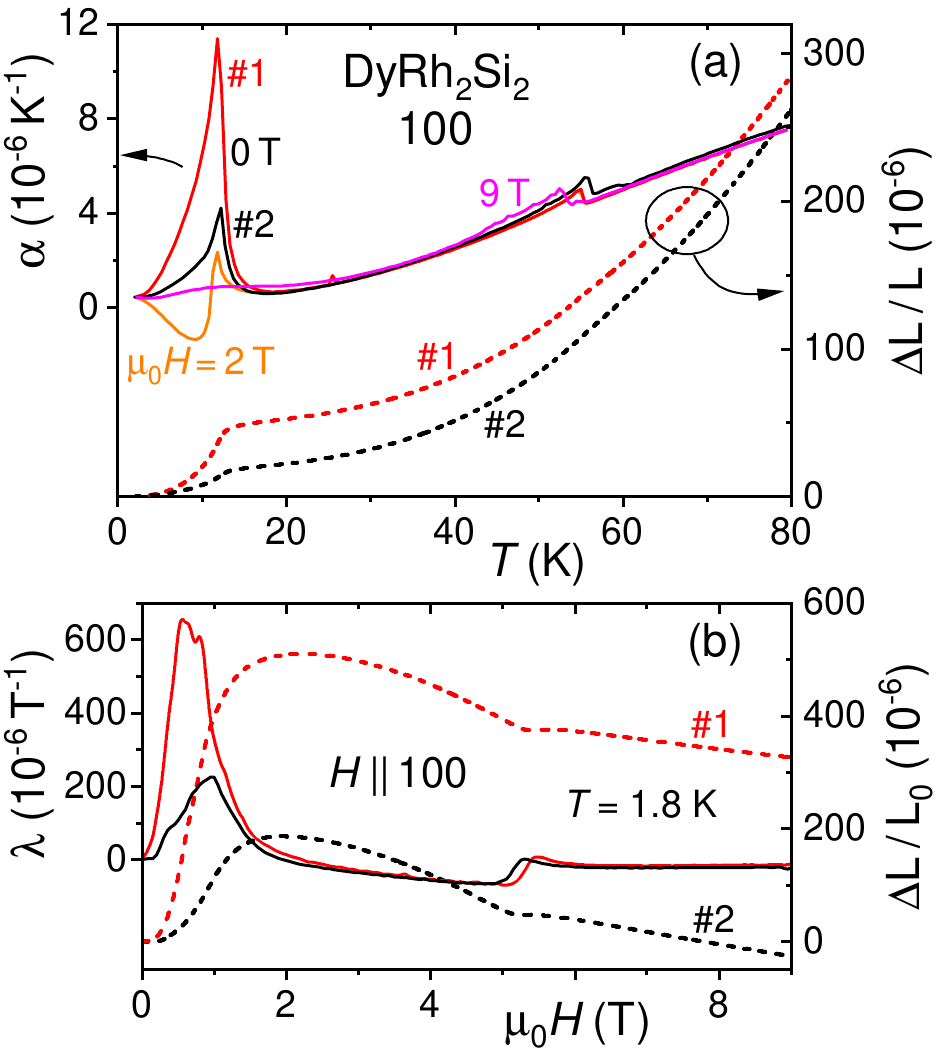}
\caption{(a) Two measurements of thermal expansion $\Delta {\rm L}_{100}(T)
/$L$_{100}$ (dashed lines, right scale) on a single crystal DyRh$_2$Si$_2$,
along with the thermal expansion coefficient $\alpha (T)$. Also included are
examples of $\alpha (T,H)$ in applied magnetic fields for run \#1. (b)
Magnetostriction $\Delta {\rm L}(H) / $L$_0$ at $T =$ 1.8~K with magnetic
fields applied along the (100) direction (dashed lines, up-sweep). The
magnetostriction coefficients $\lambda$ (left scale) vary considerably in
magnitude for $\mu_0 H \lesssim$ 1~T.}   \label{Dy-differ}
\end{figure}

Single crystals DyRh$_2$Si$_2$ exhibit a preferred natural growth edge along
the $[110]$ crystallographic direction \cite{kli20}. Therefore, a suitable
sample for measurements of thermal expansion along the (100) direction needed
to be searched for. Some results for two different runs are presented in Fig.\
\ref{Dy-differ}(a). Between these runs (denoted as \#1 and \#2) the single
crystal was mounted afresh inside our measurement cell. As can clearly be seen,
the two runs yielded quantitatively different results albeit the transitions
were always observed at very similar temperatures (or magnetic fields) and
agree well with the results from other measurements (see below). The same holds
for additional runs as well as measurements along the (110) and (001)
crystallographic directions (not shown here). This indicates that our
measurements are genuine. As in case of HoRh$_2$Si$_2$, we may speculate that
domain effects play a role in these differences but other influences cannot be
ruled out at present. We therefore restrict ourself to a discussion of the
transition temperatures (and transition fields in case of magnetostriction) in
the following.

The magnetic behavior of DyRh$_2$Si$_2$ is qualitatively similar to the one
observed for HoRh$_2$Si$_2$: The susceptibility exhibits a sharp peak at
$T_{\rm N} =$ 55~K if measured along (001), while $\chi(T)$ peaks at $T_1 =$
12~K for (100) and (110) \cite{kli23}. For both compounds, the magnetic moments
align along the $c$-direction below $T_{\rm N}$ \cite{fel83,mel84} but tilt
away from $c$ for $T < T_1$. For DyRh$_2$Si$_2$, Weiss temperatures of
$\theta_{ab} \sim -30$~K and $\theta_c \sim$ 36.5~K were reported \cite{kli23},
in line with its larger $T_{\rm N}$ compared to the Ho-compound. Also, about
twice as large magnetic fields (4.0~T and 8.2~T, \cite{kli23}) are required for
the step-like magnetization behavior in DyRh$_2$Si$_2$ with $\vec{H} \parallel
(001)$, which is otherwise very similar to the one shown in Fig.\ \ref{magn}.
There is no indication for any other transition in DyRh$_2$Si$_2$, i.e., there
appears to be no counterpart to $T_2$ seen in HoRh$_2$Si$_2$.

Included in Fig.\ \ref{Dy-differ} are examples of results for $\alpha (T,H)$
obtained at applied magnetic fields, here $\mu_0 H =$ 2~T and 9~T. For the
latter, $T_{\rm N}$ is reduced to $\sim$53~K, while the transition at $T_1$ is
largely suppressed to a faint, broad crossover. At such high in-plane fields,
the magnetic moments rotated toward the field direction regardless of the
presence/absence of any tilting away from the $c$ direction due to the CEF.

\begin{figure}[t]
\centering
\includegraphics[width=7.4cm]{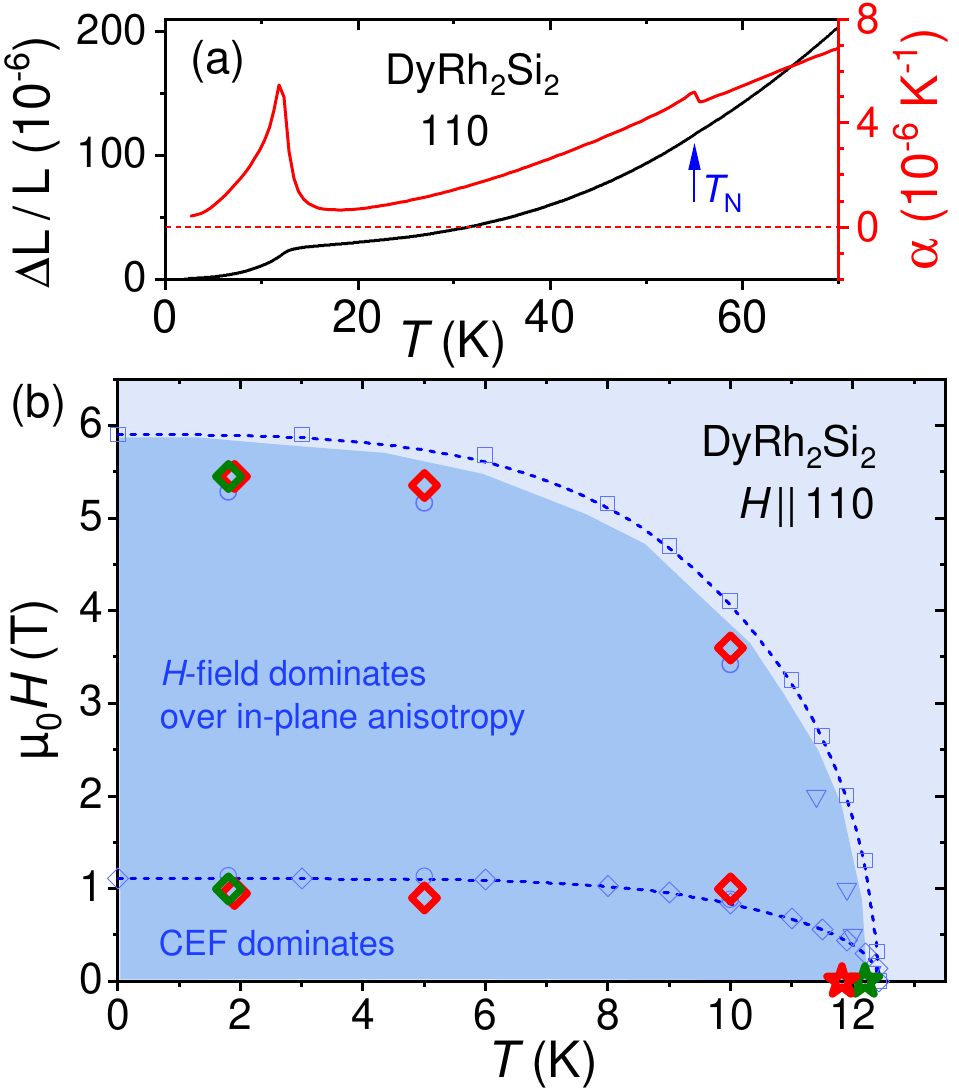}
\caption{(a) Example of a thermal expansion measurement along (110). (b)
Low temperature--magnetic field ($T$--$H$)  phase diagram of DyRh$_2$Si$_2$
with $\vec{H} \parallel (110)$. All blue (faint) data points, lines and
shadings were taken from \cite{kli23}. Our thermal expansion (stars) and
magnetostriction (diamonds) results agree well with the reported data. Green
and red colors mark results from two different samples.}
\label{PD110}  \end{figure}

An example of a thermal expansion measurement along the (110) direction is
presented in Fig.\ \ref{PD110}(a) which is, not surprisingly, very similar to
the (100) direction. The results of our measurements for $\vec{H} \parallel
(110)$ are summarized in the low temperature--magnetic field phase diagram
Fig.\ \ref{PD110}(b). All blue portions were taken from \cite{kli23}.
Obviously, the thermal expansion (stars) and magnetostriction (diamonds)
results agree nicely with these reported data. As mentioned earlier, results
obtained on different samples (marked by red and green color) also agree well
and hence, the quantitative differences discussed above are not caused by
sample dependencies. Note that the temperature range of this phase diagram is
well within the antiferromagnetic order, $T \ll T_{\rm N}$. As mentioned, below
$T_1 =$ 12~K, the magnetic moments start to tilt away from the crystallographic
$c$ direction. Moreover, the (small) magnetocrystalline anisotropy within the
basal plane gives rise to an additional feature in the phase diagram depending
on whether the CEF-derived anisotropy or the applied magnetic field dominate
energetically. In consequence, one may expect a crossover, rather than a
transition, which would explain the broad feature at $\mu_0 H <$ 1.5~T observed
in $\lambda$ for $\vec{H} \parallel (110)$ [similar to the data shown in Fig.\
\ref{Dy-differ}(b) for $\vec{H} \parallel (100)$].

\begin{figure}[t]
\centering
\includegraphics[width=8.0cm]{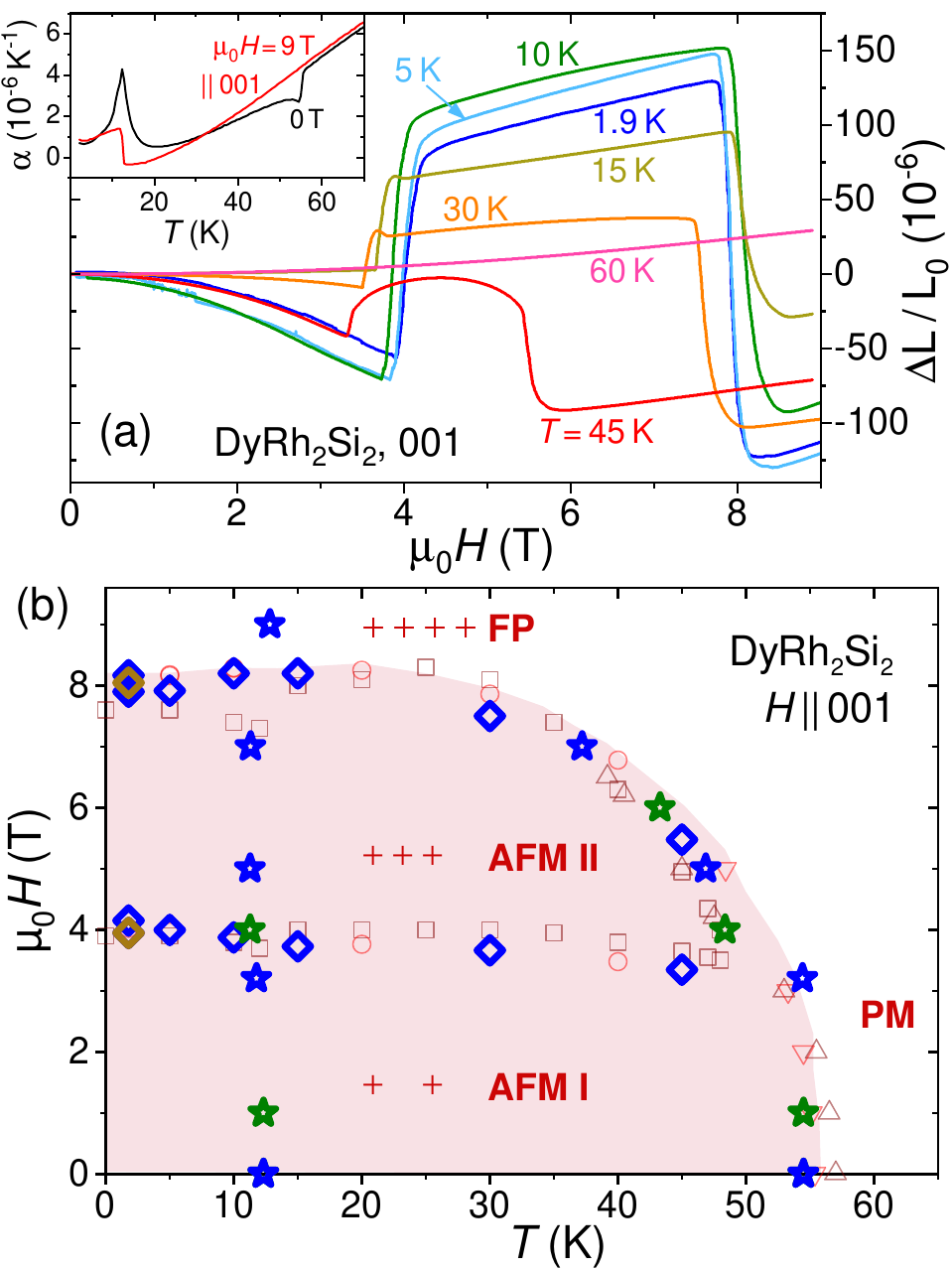}
\caption{(a) Exemplary magnetostriction $\Delta {\rm L}(H,T) / $L$_0$ of
DyRh$_2$Si$_2$ at several temperatures 1.9~K $\leq T \leq$ 60~K with magnetic
field applied $\vec{H}$ along the (001) direction. Inset: $\alpha(T,H)$
measured at 0 and 9~T for the same setup as in (a). (b) $T$--$H$ phase diagram
for $\vec{H} \parallel (110)$. All reddish (faint) data points, shading and
notes were taken from \cite{kli23}. The results of thermal expansion and
magnetostriction measurements are presented by  stars and diamonds,
respectively. Differently colored symbols (blue, green, dark yellow) mark
results from different runs on two samples. The alignment of the magnetic
moments is indicated (FP -- field polarized, PM -- paramagnetic).}
\label{PD001}  \end{figure}

An example of magnetostriction measurements with $\vec{H} \parallel (001)$ at
several temperatures is shown in Fig.\ \ref{PD001}(a). At $T =$ 60~K, i.e.\ for
$T > T_{\rm N}$, only a very small and featureless field dependence of
$\Delta {\rm L}(H) / $L$_0$ is seen. In consequence, we can safely assume
that the step-like features observed in ${\rm L}(H,T < T_{\rm N}) / $L$_0$ are
linked to the antiferromagnetic spin alignment. Plotting all the transitions
observed in the magnetostriction measurements in a $H$--$T$ phase diagram
results in Fig.\ \ref{PD001}(b). Here, our data are overlayed onto the
respective phase diagram as published in \cite{kli23} (red data points, shading
and labels). The excellent agreement in Fig.\ \ref{PD001}(b) indicates that the
two jumps observed in magnetostriction, Fig.\ \ref{PD001}(a), are related to
two consecutive spin-flip transitions from a +$-$+$-$ state (marked AFM I) to
+++$-$ (AFM II) beyond about 4~T, and finally a field-polarized ++++ state
(FP).

Our thermal expansion measurements [stars in Fig.\ \ref{PD001}(b)] conducted at
different constant fields $H$ reveal the presence of another transition which
was not included in the earlier phase diagram \cite{kli23}. As it is observed
at $T \sim $ 12~K, it is likely related to $T_1$, i.e. the temperature at which
the magnetic moments tilt away from the crystallographic $c$ direction. This
tilting is a consequence of the strong anisotropy (i.e. large related CEF
parameters \cite{kli23,tom89,tak92}) along the $c$ direction and hence, it is
not surprising to find this transition to be nearly independent of $H$, and
even in the field-polarized state at $\mu_0 H =$ 9~T (the maximum field of our
PPMS). Interestingly, despite the almost twice as high $T_{\rm N}$ of
DyRh$_2$Si$_2$ compared to HoRh$_2$Si$_2$, the temperatures $T_1$ are nearly
the same (12~K and 11.7~K, respectively). Moreover, the tilting angle of
$\sim\! 25^{\circ}$ at $T =$ 4.2~K in DyRh$_2$Si$_2$ \cite{tom89} is similar
to the one in HoRh$_2$Si$_2$. The observation of a clear jump in $\alpha(T,
\mu_0 H\!=\! 9\,{\rm T})$ at $T_1 =$ 12~K for $\vec{H} \parallel (001)$, but
only a broad hump for $\vec{H} \parallel (100)$, Fig.\ \ref{Dy-differ}(a), may
again be related to the formation of magnetic domains, but possibly also to the
(already mentioned) oblate shape of the 4$f$ electron density \cite{rin11}.

\section{Conclusion}
The compounds DyRh$_2$Si$_2$ and HoRh$_2$Si$_2$ share sizeable effects in
thermal expansion. In fact, they both exhibit a canting of the magnetic moments
away from the crystallographic $c$ direction upon cooling to temperatures below
about 12~K which are reflected in positive peaks of $\alpha (T)$ for all main
crystallographic directions. We attribute these similarities to similar CEF
effects experienced by the rare earths in both compounds as well as to similar
4$f$ electron densities of the Dy$^{3+}$ and Ho$^{3+}$ ion. The latter may also
serve as an explanation for the observed opposite changes in $\alpha (T)$
depending on whether it is measured parallel or perpendicular to the $c$
direction of the tetragonal lattice, Figs.\ \ref{Ho-TE}, \ref{Dy-differ}(a) and
inset of \ref{PD001}(a). Unfortunately, a quantitative analysis beyond this
qualitative comparison has proven difficult due to differences in the magnitude
of both $\alpha$ and $\lambda$ for differently mounted samples, particularly in
the case of DyRh$_{2}$Si$_2$. We attribute these differences to magnetic domain
effects in these samples. This assumption is supported by the fact that in all
measurements of $\alpha (T)$ and $\lambda (H)$ highly consistent transition
temperatures or transition fields, respectively, were observed.

One difference between the two compounds is the appearance of a second
transition temperature $T_2$ close to, but distinct from, $T_{\rm N}$ in
HoRh$_{2}$Si$_2$. This transition is seen in numerous properties: $\chi(T)$,
$C_p(T)$ and $\alpha (T)$. Neutron diffraction measurements indicated a strong
change of intensity of the magnetic 100 reflection at (27$\pm$1)~K
\cite{sla83}. This observation may point to a change in the magnetic structure
of HoRh$_{2}$Si$_2$ at $T_2$, an assessment in line with an analysis of the
magnetic Gr\"{u}neisen ratio.

\section{Acknowledgments}
KK and CK acknowledge funding by the Deutsche Forschungsgemeinschaft (DFG,
German Research Foundation) via SFB/TRR 288 (422213477, Project No.\ A03).

\end{document}